\documentclass[twocolumn,prc,superscriptaddress,unsortedaddress,showemail,showkeys,showpacs,preprintnumbers,amsmath,amssymb]{revtex4}
\usepackage{mathrsfs}

\usepackage{graphicx}
\usepackage{dcolumn}
\usepackage{bm}

\def\beq{\begin{equation}}
\def\eeq{\end{equation}}
\def\bea{\begin{eqnarray}}
\def\eea{\end{eqnarray}}

\def\fun#1#2{\lower3.6pt\vbox{\baselineskip0pt\lineskip.9pt
  \ialign{$\mathsurround=0pt#1\hfil##\hfil$\crcr#2\crcr\sim\crcr}}}

\begin{document}

\preprint{}

\title{Correlation between muonic levels and nuclear structure in muonic atoms }
\author{J. M. Dong}
\affiliation{Research Center for Nuclear Science and Technology,
Lanzhou University and Institute of Modern Physics of CAS, Lanzhou
730000, China} \affiliation{Institute of Modern Physics, Chinese
Academy of Sciences, Lanzhou 730000, China}\affiliation{Graduate
University of Chinese Academy of Sciences, Beijing 100049, China}
\affiliation{School of Nuclear Science and Technology, Lanzhou
University, Lanzhou 730000, China} \affiliation{China Institute of
Atomic Energy, P. O. Box 275(18), Beijing 102413, China}
\author{W. Zuo}\affiliation{Research Center for Nuclear Science and Technology, Lanzhou University and
Institute of Modern Physics of CAS, Lanzhou 730000, China}
\affiliation{Institute of Modern Physics, Chinese Academy of
Sciences, Lanzhou 730000, China} \affiliation{School of Nuclear
Science and Technology, Lanzhou University, Lanzhou 730000, China}
\author{H. F. Zhang}
\affiliation{School of Nuclear Science and Technology, Lanzhou
University, Lanzhou 730000, China}
\author{W. Scheid}
\affiliation{Institute for Theoretical Physics,
Justus-Liebig-University, D-35392 Giessen, Germany}
\author{J. Z. Gu}
\affiliation{China Institute of Atomic Energy, P. O. Box 275(18),
Beijing 102413, China}
\author{Y. Z. Wang}
\affiliation{China Institute of Atomic Energy, P. O. Box 275(18),
Beijing 102413, China}

\begin{abstract}
 A method that deals with the nucleons and the muon unitedly is
employed to investigate the muonic lead, with which the correlation
between the muon and nucleus can be studied distinctly. A
\textquotedblleft kink\textquotedblright appears in the muonic
isotope shift at a neutron magic number where the nuclear shell
structure plays a key role. This behavior may have very important
implications for the experimentally probing the shell structure of
the nuclei far away from the $\beta$-stable line. We investigate the
variations of the nuclear structure due to the interaction with the
muon in the muonic atom and find that the nuclear structure remains
basically unaltered. Therefore, the muon is a clean and reliable
probe for studying the nuclear structure. In addition, a correction
that the muon-induced slight change in the proton density
distribution in turn shifts the muonic levels is investigated. This
correction to muonic level is as important as the Lamb shift and
high order vacuum polarization correction, but is larger than
anomalous magnetic moment and electron shielding correction.

\end{abstract}



\maketitle

A muonic atom is an atom in which one of the electrons has been
replaced by a negatively charged muon. Because of the very large
mass of a muon compared with a electron and the correspondingly
small Bohr radius, the muonic wave function has a large overlap with
the nucleus. X-ray transition energies in muonic atoms are strongly
affected by the size of the nuclei, and can be used efficiently to
determine the nuclear charge distribution~\cite{WHEE,TTA}. A
detailed introduction about muonic atoms can be found in
Ref.~\cite{RMP}. Moreover, the muonic atom tends to play an
important role in investigation on other subjects. A recent study
demonstrated that muonic atoms in strong laser fields can be used to
dynamically gain structure information on nuclear ground
states~\cite{CC}. Masafumi Koike {\it et al.} proposed a new process
of $\mu ^{-}e^{-}\longrightarrow e^{-}e^{-}$ in a muonic atom for a
quest of charged lepton flavor violation (This violation is known to
be one of the important rare processes to search for new physics
beyond the standard model) in consideration of the fact that this
process in a muonic atom has various significant
advantages~\cite{Vio}. An attractive means to improve the accuracy
in the measurement of proton root-mean square radius is provided by
muonic hydrogen~\cite{size}. In addition, some potential synergies
of combining muons with radioactive nuclei may become a new tool to
be used at future RIB facilities, as suggested in Ref.~\cite{AAA}.
Therefore, investigation on muonic atom is meaningful and not
limited to atomic physics.

We extend the relativistic mean field (RMF) approach in this Letter
to include the negatively charged muon, and then the correlation
between the muon and nucleus, namely the effects of the nuclear
structure on the muonic spectra as well as the influence of the muon
on the nuclear structure, can be investigated distinctly. Within
this correlation, we may find some new approach to probe the nuclear
structure. Only one muon captured by the nucleus is discussed here.
Because the motion of the bound muon is relativistic for high-$Z$
atoms, it is necessary to treat it in relativistic framework.
Nowadays the RMF theory has became a standard tool in low energy
nuclear structure ~\cite{RMF1,RMF2,RMF3} and the interacting
Lagrangian density taking into account the muonic field is given by
\begin{eqnarray}
\mathscr{L} &=&\overline{\psi }(i\gamma ^{\nu }\partial _{\nu }-M)\psi +\frac{1}{2}%
\partial _{\nu }\sigma \partial ^{\nu }\sigma   \notag \\
&&-(\frac{1}{2}m_{\sigma }^{2}\sigma ^{2}+\frac{1}{3}g_{2}\sigma ^{3}+\frac{1%
}{4}g_{3}\sigma ^{4})-g_{\sigma }\overline{\psi }\psi \sigma   \notag \\
&&-\frac{1}{4}\Omega _{\nu \lambda }\Omega ^{\nu \lambda }+\frac{1}{2}%
m_{\omega }^{2}\omega _{\nu }\omega ^{\nu }-g_{\omega }\overline{\psi }%
\gamma ^{\nu }\psi \omega _{\nu }  \notag \\
&&-\frac{1}{4}{\bm R}_{\nu \lambda }\cdot {\bm R}^{\nu \lambda
}+\frac{1}{2}m_{\rho }^{2}{\bm \rho} _{\nu }\cdot {\bm \rho} ^{\nu
}-g_{\rho }\overline{\psi }\gamma ^{\nu
}{\bm \tau} \cdot {\bm \rho _{\nu}} \psi   \notag \\
&&-\frac{1}{4}F_{\nu \lambda }F^{\nu \lambda }-e\overline{\psi
}\gamma ^{\nu
}\frac{1+\tau _{3}}{2}\psi A_{\nu }\nonumber \\
&&+\overline{\psi _{\mu }}(i\gamma ^{\nu }\partial _{\nu }-m_{\mu
})\psi _{\mu }-(-e)\overline{\psi _{\mu }}\gamma ^{\nu }\psi _{\mu
}A_{\nu }^{^{\prime }}.
\end{eqnarray}
$M$, $m_{\sigma }$, $m_{\omega }$ and $m_{\rho }$ are the nucleon-,
the $\sigma $-, the $\omega$- and the $\rho $-meson masses,
respectively. The muon mass of $m_{\mu }=105.6583668$ MeV is taken
from Ref.~\cite{mass}. The nucleon field $\psi $ interacts with the
$\sigma ,\omega ,\rho $ meson fields $\sigma ,\omega _{\nu },\rho
_{\nu }$ and with the photon field $A_{\nu }$. The muonic field
$\psi _{\mu }$ interacts with the photon field $A_{\nu }^{^{\prime
}}$ and $A_{\nu }^{^{\prime }}$ excludes the photon field produced
by the muon itself. The field tensors for the vector meson are given
as $\Omega _{\nu\lambda }=\partial _{\nu }\omega _{\lambda
}-\partial _{\lambda }\omega _{\nu }$
 and by similar expression for $\rho $ meson and the photon. The
 self-coupling terms with coupling constants $g_{2}$ and $g_{3}$ for
 the $\sigma $ meson are introduced which turned out to be
 crucial~\cite{BB}. Since the muon couples only electromagnetically to
 nucleons, no additional parameters need to be introduced, and no
 readjustment of the present parameters is needed. Varying the effective Lagrangian, one can obtain the Dirac
 equation for nucleons and muons, and the Klein-Gordon equations for
 mesons. The calculations are performed in
coordinate space using a mesh size of 0.01 fm and different box
sizes for different muonic states, and the paring correlations are
accounted in the BCS formalism with an energy gap $\Delta$ obtained
from the observed odd even mass differences for an open shell and
$\Delta=0$ for a closed shell. The NL3 and NLSH parameter sets are
employed here. The NL3 parameter set has been used with enormous
success in the description of a variety of ground-state properties
of spherical, deformed and exotic nuclei~\cite{GLJK,BGT}, and NLSH
is also a successful parameter set~\cite{NLSH}. The nucleons and
muon are treated in the unified framework without any adjustable
parameter and the finite size effect of the nucleus is automatically
included so that both the muonic atom structure and nuclear
structure can be investigated simultaneously. In contrast to earlier
works about the muonic atoms based on Migdal theory~\cite{NN1,NN2},
the present method is relativistic. Also, our approach is more
microscopic compared with the method that directly solves the Dirac
equation using the two or three-parameter Fermi-type distribution of
the nuclear charge~\cite{EXP,HJM}. In fact, it was a common practice
to fit the parameter sets for this nuclear charge distributions with
the help of the experimental muonic levels~\cite{EXP}. In the RMF
theory, the nucleons are treated as point particles, which is a
drawback of RMF theory. As a matter of fact, the finite size effects
of the nucleons cause some influence on the nuclear properties. The
muonic spectrum is affected by the finite size effects of the
protons because the distributions of charges differ slightly from
those of the protons. But in the self-consistent method, the proton
density distributions are used as sources rather than the charge
distributions. In our calculations, the Coulomb potential that the
muon feels excludes the potential produced by the muon itself. In
other words, the muon only interacts with the electrostatic
potential generated by the protons. In fact, the mean field that the
muon feels is a central potential field for a spherical nucleus, and
the corresponding muonic energy $E_{\mu0}$ is obtained by applying
the RMF approach discussed above.

\begin{table*}
\label{table1} \caption{The muonic spectrum obtained with the RMF
method including the lowest order vacuum polarization correction for
the muonic $^{204,206,208}$Pb. The mean field part $E_{\mu0}$ and
the lowest order vacuum polarization correction $\Delta \varepsilon
_{\mu }$ are listed and the muonic levels are given by
$E_{\mu}=E_{\mu0}+\Delta \varepsilon _{\mu}$. The experimental
data~\cite{EXP} have been listed for comparison. All energies are in
units of keV. }
\begin{ruledtabular}
\begin{tabular}{lllllllll}
A & orbit & $E_{\mu0}$(NL3)  &$\Delta \varepsilon _{\mu }$(NL3) & $E_{\mu0}$(NLSH)  &$\Delta \varepsilon _{\mu }$(NLSH) & $E_{\mu }{\text{(NL3)}}$&$E_{\mu }{\text{(NLSH)}}$ &  $E_{\mu }{\text{(Expt.)}}$ \\
\hline
204 &1$s_{1/2}$ & -10549.41 & -66.35  & -10561.24   & -66.47  &-10615.76 &-10627.72      &$-10614.88\pm0.41$ \\
204& 2$p_{1/2}$ &-4781.17 & -31.26    &  -4783.41  & -31.30  &-4812.43   &-4814.71      & $-4818.62\pm0.08$ \\
204& 2$p_{3/2}$ &-4597.97  &-28.70    &  -4599.67  &-28.73 &  -4626.67 & -4628.40      & $-4632.72\pm0.06$ \\
204& 2$s_{1/2}$ &-3583.87  &-18.26    &  -3585.90  &-18.28   &-3602.13 &-3604.18      & $-3604.14\pm0.26$ \\
204& 3$d_{3/2}$ & -2161.81  &-9.40    & -2161.86   &-9.40  &-2171.21 &-2171.26      & $-2173.03\pm0.05$\\
204& 3$p_{3/2}$ & -2082.71 & -9.20    & -2083.28  &-9.21  &-2091.90     &-2092.49       &$-2092.41\pm0.33$\\
206 &1$s_{1/2}$ & -10540.54  &-66.25  & -10551.51  &-66.37   &-10606.80   & -10617.87      &$-10604.34\pm0.41$\\
206& 2$p_{1/2}$ &  -4779.84  &-31.23  & -4781.97  & -31.27      & -4811.07 &-4813.24     & $-4817.19\pm0.06$ \\
206& 2$p_{3/2}$ &  -4597.03  &-28.68  & -4598.65  & -28.71     & -4625.71   &-4627.37      & $-4631.50\pm0.05$\\
206& 2$s_{1/2}$ &  -3582.23  & -18.24 &  -3584.08   &-18.26    &-3600.47  &-3602.34      & $-3601.54\pm0.34$\\
206& 3$d_{3/2}$  & -2161.78  &-9.40   & -2161.84 & -9.40      & -2171.18   & -2171.23     & $-2173.02\pm0.05$\\
206& 3$p_{3/2}$ & -2082.38  &-9.19    &  -2082.93   &-9.20   & -2091.57  & -2092.13       &$-2092.01\pm0.39$\\
208 &1$s_{1/2}$ & -10532.13   &-66.16 &  -10542.23  &-66.26  &-10598.29       &-10608.49      &$-10593.13\pm0.41$ \\
208& 2$p_{1/2}$ & -4778.56   & -31.21 & -4780.59  & -31.25  & -4809.77  & -4811.83     & $-4815.14\pm0.06$ \\
208& 2$p_{3/2}$ & -4596.13  &-28.66   &  -4597.68   &-28.69  & -4624.79 &-4626.37    & $-4630.25\pm0.05$ \\
208& 2$s_{1/2}$ & -3580.66   &-18.22  &   -3582.35   &-18.24  &-3598.88  & -3600.58      & $-3599.75\pm0.17$ \\
208& 3$d_{3/2}$ &  -2161.76  & -9.40  &  -2161.81  & -9.40  & -2171.15  &-2171.21      & $-2173.01\pm0.05$\\
208& 3$p_{3/2}$ & -2082.07  &-9.19    &  -2082.59  & -9.20  &-2091.25   & -2091.79       &$-2092.20\pm0.32$\\

\end{tabular}
\end{ruledtabular}
\end{table*}

\begin{figure}[htbp]
\begin{center}
\includegraphics[width=0.4\textwidth]{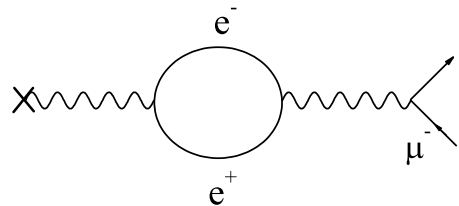}
\caption{Lowest order vacuum polarization diagram.}
\end{center}
\end{figure}

Once the Dirac wave functions of the muon are obtained with the RMF
method, the lowest order vacuum polarization correction $\Delta
\varepsilon_{\mu }$, as shown in Fig. 1, can be calculated
additionally. The details on this lowest order vacuum polarization
correction can be found in Ref.~\cite{HJM}. The ultimate calculated
muonic levels $E_{\mu}=E_{\mu0}+\Delta \varepsilon _{\mu}$ in the
muon-Pb systems are presented in Table I. As can be seen, the
results of the RMF method taking into account lowest order vacuum
polarization corrections agree with the experimental data suggesting
the effectiveness of the RMF to describe the muonic atoms. The
lowest order vacuum polarization correction is relatively large for
low energy levels, but the correction become smaller for high energy
levels. The nuclear polarization that arises the muon-nucleus
electrostatic interaction excites the nucleus into virtual excited
states, is partly taken into account due to the self-consistent
calculations, as will be shown below. Other corrections such as high
order vacuum polarization correction, Lamb shift, anomalous magnetic
moment, electron shielding and static hyperfine structure
interaction, contribute little to the binding energy of the muon
atoms compared with the lowest order vacuum polarization
correction~\cite{BS} and these effects are inessential for the
following discussions, so they are not considered here. Note that
the uncertainties of muonic levels especially 1$s_{1/2}$ levels in
heavy nuclei are mainly ascribed to the nuclear structure, i.e., the
proton density distribution. The agreement between the calculated
and experimental muonic levels indicates the reliability of the
proton distribution from the RMF calculations. The 1$s_{1/2}$ levels
with NL3 are in better agreement with the experimental data than
those obtained with NLSH, possibly because that NL3 can provide a
more accurate proton density distribution over the parameter set
NLSH. The parametrization NL3 is more excellent because it cures
some deficiencies of the NLSH to some extent, as suggested in
Ref.~\cite{GLJK}. In addition, the mean speed of the muon in
1$s_{1/2}$ orbit for the muon-$^{208}$Pb system is $v=0.31c$
according to our calculations, indicating the importance of
relativistic effect. Thus the treating of the muon and nucleons in a
unified relativistic framework is necessary.

\begin{figure}[htbp]
\begin{center}
\includegraphics[width=0.5\textwidth]{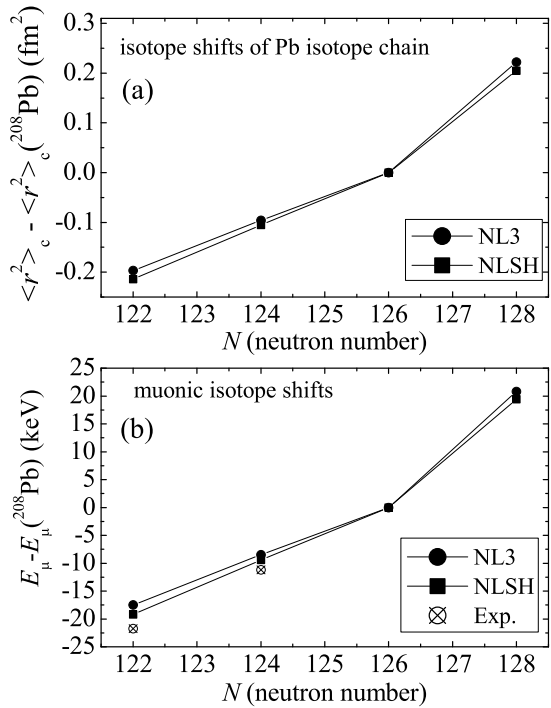}
\caption{(a) Nuclear isotope shifts of the Pb isotope chain and (b)
muonic isotope shifts of 1$s_{1/2}$ state in muonic Pb. The
$^{208}$Pb is taken as a reference nucleus and the experimental data
are taken from Ref.~\cite{EXP}. }
\end{center}
\end{figure}

In Fig. 2(a), the nuclear isotope shifts $\langle r^{2}\rangle
_{c}-\langle r^{2}\rangle _{c}(^{208}$Pb$)$ for the Pb isotope chain
are plotted as a function of neutron number, taking the nucleus
$^{208}$Pb as a reference. With the increase of the neutron number
$N$, the charge radius squared changes only slightly until it
reaches the magic neutron number $N=126$, but it increases more
rapidly with the neutron number $N$ goes beyond $N=126$. In other
words, a \textquotedblleft kink\textquotedblright in the nuclear
isotope shift is found at $N=126$. We would like to mention that the
non-relativistic calculations have not succeeded in offering this
fact which originates from the nuclear shell effect. The muonic
isotope shift (isotope shift of atomic spectrum) which is the shift
of the muonic energy with increasing neutron number, includes the
field shift coming from the change of spacial distribution of the
charges in the nucleus and mass shift originating from the change of
the mass of the nucleus. For a heavy nucleus, the field shift is
major and much larger than the mass shift. In heavy elements such as
lead, the mass effect is negligible and the field effect roughly
accounts for the observed shifts. Fig. 2(b) displays the muonic
isotope shift taking the muonic energy in $^{208}$Pb as a reference.
Hence computational errors along with various corrections cancel to
a high degree. This muonic isotope shift can be attributed to the
nuclear isotope shift. The larger the nuclear charge radius is, the
wider the charge distributes since the total charge is invariant,
and hence a high muonic energy. A \textquotedblleft
kink\textquotedblright in the muonic isotope shift is also found at
$N=126$, which stems from the neutron shell effect in the final
analysis. For other isotope chains, the \textquotedblleft
kink\textquotedblright can be also found at magic number, and we do
not discuss them in detail here. This is certainly a good news for
the nuclear structure study. The \textquotedblleft
kink\textquotedblright in the muonic isotope shifts may be taken as
a useful signal to experimentally probe the shell structure of the
nuclei far away from the $\beta$-stable line for which novel nuclear
structure effects may exist. This is one of the main conclusions we
draw in this work. Strasser {\it {et al.}} has proposed the cold
hydrogen film method to extend muonic atom spectroscopy to the use
of nuclear beams including radioactive isotope beams to produce
radioactive muonic atom in the future~\cite{STR}. This would allow
studies of unstable nuclei by means of the muonic X-ray method at
facilities where both $\mu^{-}$ and radioactive isotope beams would
be available. The electron scattering is not going to be easy to be
applied for exotic nuclei with very low beam intensities when one
measures the scattering cross sections to obtain the knowledge about
the nuclear structure. But for muonic atoms, one measures the X-ray
energies, which is much easier and X-ray energy can be measured with
a high accuracy. For the other approach to probe the nuclear shell
structure, the first excited 2$^{+}$ state, can be carried out using
Coulomb excitation~\cite{CE0} for these exotic nuclei. Therefore, as
two independent methods, the approaches of Coulomb excitation and
muonic atom spectroscopy can complement each other.

\begin{table}[h]
\label{table2} \caption{The correction for muonic 1$s_{1/2}$ orbit
due to the muon-induced slight change of the proton density
distributions in muonic Pb.}
\begin{ruledtabular}
\begin{tabular}{llllllll}
Nucleus& NL3 & NLSH\\
\hline
$^{204}$Pb& -1.5 keV &-1.2 keV \\
$^{206}$Pb& -1.5 keV &-1.2 keV\\
$^{208}$Pb& -1.5 keV &-1.2 keV\\
$^{210}$Pb& -1.5 keV &-1.2 keV\\
\end{tabular}
\end{ruledtabular}
\end{table}

Apart from obtaining information on the muonic spectrum, the
influence of the muon on nuclear structure can also be investigated.
It is found that the nuclear structure of the Pb isotope remains
basically unchanged in the presence of the muon in 1$s_{1/2}$ orbit.
The single particle level spacing of neutrons and protons is altered
by only several or some dozen keV. The nucleon densities are reduced
by about $0.1\%$ at the edge of the nucleus ($\sim $7 fm) but
enhanced in the interior of nucleus by about $0.1\%$. In other
words, the protons and neutrons move quite slightly towards the core
as a result of the bound muon, and the nuclear rms radii are reduced
only by about $0.02\%$. For the calculation of these changes in
value, the systematic computational errors here could be canceled to
a large extent, so that the results are reliable. The bound muon in
other orbits affect the nuclear structure much more weakly than that
in the 1$s_{1/2}$ orbit. The muon thus can be taken as a clean and
reliable probe to extract information on nuclear structure since
this probe can hardly change the nuclear structure. It is reliable
to yield information on nuclear charge distributions and rms radii
through muonic X-ray in experiments. That is to say, the previous
experiments on the nuclear charge distributions and charge rms radii
within the muonic atoms were theoretically confirmed to be
reasonable and reliable. Moreover, the fact that the nuclear
structure is only slightly changed by the muon implies that the
linear response theory used in Ref.~\cite{NN1,NN2} is an excellent
approximation. Although the change of the proton density
distribution due to the bound muon is quite slight, it in turn
lowers the muonic levels especially for 1$s_{1/2}$ orbit, which is a
part of nuclear polarization since only virtual excitation of
$J^{\pi}=0^{+}$ states is taken into account here. This correction
for muonic Pb is presented in Table II, which is more important than
the anomalous magnetic moment and electron shielding, and can be
compared to Lamb shift and high order vacuum polarization
corrections (For the 1$s_{1/2}$ orbit in muonic $^{206}$Pb, the
corrections to muonic energy levels from anomalous magnetic moment,
Lamb shift and high order vacuum polarization are 0.445
keV~\cite{BS}, 2.302 keV ~\cite{BS} and -1.75 keV~\cite{BS1},
respectively. The electron shielding correction is even less, for
example, -4.6 eV for the 1$s_{1/2}$ orbit in muonic
$^{209}$Bi~\cite{BLT}). However, this effect tended to be neglected
in many previous investigations which regarded the charge
distribution as an invariant when a muon is captured. Yet it is not
easy to be studied with the usual method but can be automatically
included in our approach since the muon and nucleons are treated in
a unified theoretical framework. Attention should be paid attention
to this correction since the experimental measurements on muonic
levels can achieve this precision.

In summary, the muonic atoms have been investigated in the framework
of many-body theory. The main conclusions are summarized as follows:
(1) The relativistic mean field theory for nuclei has been extended
to investigate the muonic atoms of lead isotopes by taking the
nucleons and a bound muon as a system and the theoretical muonic
levels agree with the experimental data. Therefore, a direct link
has been established between the nuclear physics and atomic physics.
(2) The muonic isotopes shift stems from nuclear isotope shift for
heavy nuclei. Just like the \textquotedblleft kink\textquotedblright
in nuclear isotope shift, a \textquotedblleft kink\textquotedblright
also appears in the muonic isotope shift at the neutron magic
number. Perhaps it gives us a new approach to experimentally probe
the shell structure of the nuclei far from the $\beta$-stable line
via this \textquotedblleft kink\textquotedblright in muonic isotope
shift. (3) The muon-induced changes of nuclear structure are quite
slight, which indicates the muon can be taken as a clean and
reliable probe to extract information on nuclear structure, and
theoretically confirms that the previous experiments on the nuclear
charge distributions and charge rms radii within the muonic atoms
were reasonable and reliable. (4) As a part of nuclear polarization,
the correction that the muon-induced slight change in the proton
density distribution in turn shifts the muonic levels is studied,
which should be paid attention to because the contribution of it to
muonic levels is more important than anomalous magnetic moment and
electron shielding, and can be comparable to the Lamb shift and high
order vacuum polarization correction.

J. M. Dong is thankful to Prof. Umberto Lombardo for helpful
discussions. This work is supported by the National Natural Science
Foundation of China (10875151,10575119,10975190,10947109), the Major
State Basic Research Developing Program of China under No.
2007CB815003 and 2007CB815004, the Knowledge Innovation Project
(KJCX2-EW-N01) of Chinese Academy of Sciences, CAS/SAFEA
International Partnership Program for Creative Research Teams
(CXTD-J2005-1), the Funds for Creative Research Groups of China
(Grant 11021504) and the financial support from DFG of Germany.

\end{document}